\documentclass[11pt]{article}
\usepackage{amsfonts}

\input epsf  

\newcommand{\sect}[1]{\setcounter{equation}{0}\section{#1}}
\newcommand{\subsect}[1]{\subsection{#1}}
\newcommand{\subsubsect}[1]{\subsubsection{#1}}

\def\be{\begin{equation}}
\def\ee{\end{equation}}
\def\bea{\begin{eqnarray}}
\def\eea{\end{eqnarray}}

\def\1{\'{\i}}                           
 
\def\>#1{{\bf #1}}                 
\def\d{{\rm d}}

\def\te{\theta}
\def\rr{\rho}

\def\YY{Q}

\def\jp{J_+}
\def\jm{J_-}
\def\jj{J_3}

\def\cte{\alpha}
\def\cteb{\gamma}
\def\ctec{k}
\def\otra{b}

\def\SW{\rm ISW}
\def\Stc{\rm SSW}
\def\SStc{\rm SW}
\def\kc{\rm IKC}
\def\In{\rm I}

\def\la{\lambda}
\def\pot{{\cal U}}

\begin{document}

\thispagestyle{empty}

\ 
\hfill\

\begin{center}

{\LARGE{\bf{Integrable potentials on spaces with
curvature}}}

{\LARGE{\bf{from quantum groups}}}

\bigskip

\end{center}

\bigskip 
\medskip

\begin{center}    Angel~Ballesteros$^a$,
 Francisco~J.~Herranz$^a$ and Orlando Ragnisco$^{b}$ 
\end{center}

\begin{center} {\it { 
${}^a$Departamento de F\1sica, Universidad de Burgos, Pza.\
Misael Ba\~nuelos s.n., \\ E-09001 Burgos, Spain }}\\ e-mail:
angelb@ubu.es, fjherranz@ubu.es
\end{center}

\begin{center} {\it { 
${}^b$Dipartimento di Fisica,   Universit\`a di Roma Tre and Instituto
Nazionale di Fisica Nucleare sezione di Roma Tre,  Via Vasca Navale 84,\\
I-00146 Roma, Italy}}\\ e-mail: ragnisco@fis.uniroma3.it
\end{center}

\bigskip 
\medskip

\begin{abstract}
A family of classical integrable systems defined on a deformation of the
two-dimensional sphere, hyperbolic and (anti-)de Sitter spaces is
constructed through Hamiltonians defined on the non-standard quantum
deformation of a $sl(2)$ Poisson coalgebra. All  these spaces have a
non-constant curvature   that  depends on the deformation
parameter $z$. As particular cases, the analogues of the harmonic
oscillator and Kepler--Coulomb potentials on such spaces are
proposed. Another deformed   Hamiltonian is also 
shown to provide superintegrable systems on the usual 
sphere, hyperbolic and (anti-)de Sitter spaces with a constant
curvature that exactly coincides with   $z$.
According to each specific space, the resulting potential is
interpreted as the superposition of  a central  harmonic
oscillator     with either two more
oscillators or centrifugal barriers.  The non-deformed limit
$z\to 0$ of all  these Hamiltonians can then be regarded as
the zero-curvature limit (contraction) which leads to the corresponding
(super)integrable systems on the flat Euclidean and Minkowskian spaces.

\end{abstract}

\bigskip\bigskip 

\noindent
PACS: 02.30.lk \quad 02.20.Uw

\bigskip\medskip


\newpage

\sect{Introduction}

One of the possible applications of quantum deformations of 
groups and algebras~\cite{Dr,Dri,CP,majid,Sierra} is the
 construction of classical and quantum
integrable systems  with an arbitrary number
 of degrees of freedom which was presented in~\cite{BR}. In this context, 
Poisson coalgebras (Poisson algebras
endowed with a compatible coproduct structure)  have been
shown to generate in a systematic way
certain (super)integrable classical Hamiltonian systems.  In this
construction, once a  symplectic realization of the algebra is given, the
generators of the Poisson coalgebra play the role of  dynamical
symmetries of the Hamiltonian, while   the coproduct   is used to
`propagate' the integrability to arbitrary
dimension. From this coalgebra approach,  several well-known classical
(super)integrable systems have been recovered and some integrable
deformations for them, as well as new integrable systems,   have
also  been 
obtained~\cite{BR,Deform,photon,photonb,cluster,CRMAngel}.

Recently, this
 integrability-preserving deformation procedure   has  been
used to introduce both superintegrable and integrable free motions on
two-dimensional (2D) spaces with curvature, either constant or variable,
respectively~\cite{plb}. Therefore one could expect that potential terms
can also be considered, in such a way that the coalgebra approach  should
provide (super)integrable potentials on curved spaces. The aim of the
present paper is to prove this assertion through the   construction
of some relevant Hamiltonians.

In order to make these ideas more explicit, let us   consider the
non-standard quantum deformation of
$sl(2)$~\cite{Ohn} written as a Poisson coalgebra $(sl_z(2), \Delta_z)$ with
(deformed)  Poisson brackets,  coproduct and  Casimir given by
\be 
\{\jj,\jp\}=2 \jp \cosh z\jm  \qquad 
 \{\jj,\jm\}=-2\,\frac {\sinh z\jm}{z}\qquad
\{\jm,\jp\}=4 \jj   
\label{ba}
\ee 
\be  
 \Delta_z(\jm)=  \jm \otimes 1+
1\otimes \jm \qquad 
 \Delta_z(J_i)=J_i \otimes {\rm e}^{z \jm} + {\rm e}^{-z \jm} \otimes
J_i \quad\ i=+,3  
\label{bb}
\ee 
\be
{\cal C}_z= \frac {\sinh z\jm}{z} \jp -\jj^2   
\label{bc}
\ee
where $z$ is a real deformation parameter.
A two-particle symplectic realization of (\ref{ba}) in terms of
two canonical pairs of coordinates $(q_1,q_2)$ and momenta $(p_1,p_2)$
with respect to the usual Poisson bracket
\be
\{f,g\}=\sum_{i=1}^2\left(\frac{\partial f}{\partial q_i}
\frac{\partial g}{\partial p_i}
-\frac{\partial g}{\partial q_i} 
\frac{\partial f}{\partial p_i}\right),  
\label{ag}
\ee
and that depends on two real parameters $\otra_1,\otra_2$,
reads~\cite{Deform,CRMAngel}
\be  
\begin{array}{l}
\displaystyle{ \jm=q_1^2+q_2^2\qquad\quad \jj=
\frac {\sinh z q_1^2}{z q_1^2 } \, q_1 p_1  \, {\rm e}^{z q_2^2} +
\frac {\sinh z q_2^2}{z q_2^2 }\,  q_2 p_2  \, {\rm e}^{-z q_1^2} }\\[10pt]
\displaystyle{  \jp=
\left( \frac {\sinh z q_1^2}{z q_1^2}\,  p_1^2 +\frac{z \otra_1}{\sinh
z q_1^2} \right) {\rm e}^{z q_2^2} +
\left(\frac {\sinh z q_2^2}{z q_2^2} \, p_2^2  +\frac{z \otra_2}{\sinh z
q_2^2} \right) {\rm e}^{-z q_1^2} } .
\end{array}
\label{be}
\ee 
 By substituting (\ref{be}) in
(\ref{bc}) we obtain
 the two-particle Casimir  
\bea
&& {\cal C}_z = \frac {\sinh z q_1^2
}{z q_1^2 } \,
\frac {\sinh z q_2^2}{z q_2^2} 
\left({q_1}{p_2} - {q_2}{p_1}\right)^2
{\rm e}^{-z q_1^2}{\rm e}^{z q_2^2} +(\otra_1 {\rm e}^{2z q_2^2} +\otra_2 {\rm e}^{-2z
q_1^2})\cr 
&&\qquad\qquad +\left( \otra_1\, \frac {\sinh z q_2^2}{\sinh z q_1^2}
+ \otra_2\, \frac {\sinh z q_1^2}{\sinh z q_2^2} \right) {\rm e}^{-z q_1^2}{\rm e}^{z
q_2^2}  
\label{bf}
\eea 
which Poisson-commutes with the generators (\ref{be}).   The
limit $z\rightarrow 0$ of such deformed generators leads to the well-known
symplectic realization of $sl(2)$:
\be  
\begin{array}{l}
\displaystyle{ \jm=q_1^2+q_2^2\qquad\quad \jj=
q_1 p_1 +  q_2 p_2   }\qquad 
\displaystyle{  \jp=
 p_1^2 +\frac{\otra_1}{
q_1^2} + p_2^2 +\frac{\otra_2}{
q_2^2}   } .
\end{array}
\label{beeb}
\ee

The coalgebra approach~\cite{BR}  ensures that {\em any}
smooth function  ${\cal H}_z={\cal H}_z(\jm,\jp,\jj)$
defined
on (\ref{be}) provides an integrable Hamiltonian,  for which 
${\cal C}_z$  is the constant of the motion. In this paper we
shall study  some   choices for ${\cal H}_z$ that lead to 
Hamiltonians which are quadratic in the momenta and belong to the family
 \be  
{\cal H}_z=\frac 12 \jp \, f (z\jm 
 )+\pot (z\jm )  
\label{ahaa}
\ee
where $f$ and $\pot$ are arbitrary  smooth functions such that the
   $\lim_{z\to 0}\pot(zJ_-)$ is well defined  
  and $\lim_{z\to 0}f(zJ_-)=1$. 

Therefore integrable deformations of the free motion  of a particle on
the 2D Euclidean space are obtained from (\ref{ahaa}) by setting  $b_1=b_2=0$ and
$\pot=0$. Two main
representative cases appear~\cite{plb}:

\begin{itemize}

\item The simplest {\em integrable} Hamiltonian 
\be  
\begin{array}{l}
{\cal H}_z^{\rm I}=\frac 12 \jp  
\end{array}
\label{freei}
\ee
which defines the  geodesic
motion on  a 2D Riemannian space with  metric 
\be
\d s^2=\frac {2z q_1^2}{\sinh z
q_1^2} \, {\rm e}^{-z q_2^2} \,\d q_1^2   +
 \frac {2 z q_2^2}{\sinh z q_2^2} \, {\rm e}^{z q_1^2}\, \d q_2^2    
\label{cc}
\ee
and whose
non-constant   Gaussian curvature $K$ is given  by 
\be
K(q_1,q_2)=-   z \sinh\left(z(q_1^2+q_2^2) \right)  .
\label{cd}
\ee

\item The {\em superintegrable} Hamiltonian 
\be  
\begin{array}{l}
{\cal H}_z^{\rm S}=\frac 12 \jp \,{\rm e}^{ z \jm} 
\end{array}
\label{frees}
\ee
that leads to a  Riemannian   metric of constant curvature  which
coincides with the deformation parameter, $K=z$, namely
\be
\d s^2=\frac {2z q_1^2}{\sinh z
q_1^2} \, {\rm e}^{-z q_1^2}{\rm e}^{-2 z q_2^2} \,\d q_1^2   +
 \frac {2 z q_2^2}{\sinh z q_2^2} \, {\rm e}^{-z q_2^2} \, \d q_2^2   .
\label{ec}
\ee
\end{itemize}
Consequently,  the `classical' limit  $z\to 0$   corresponds to a  
zero-curvature limit. Section 2 is devoted to the explicit solution of the
geodesic flows on all  these spaces, that complete the preliminary
description given in~\cite{plb} and include deformations of the 2D
 sphere and hyperbolic  spaces as well as of the  $(1+1)$D 
(anti-)de Sitter spacetimes.
 
  The introduction of integrable
potentials with coalgebra symmetry is then analysed by making use of the
 function $\pot$ and taking both parameters $\otra_i$ arbitrary. In fact, any $\pot$
such that
$\lim_{z\to 0}
\pot = \beta_0\jm$ can be interpreted as a deformation of the well-known
 2D Smorodinsky--Winternitz (SW)
system~\cite{Fris,Evansa,Evansb,Groschea} formed by an isotropic harmonic
oscillator   with angular frequency $\omega=\sqrt{\beta_0}$ plus two `centrifugal
terms' governed by  $\otra_1, \otra_2$:
\be
 {\cal H}^{\rm SW} = \frac 12 \left( p_1^2 + p_2^2 \right)+
\frac{\otra_1}{2 q_1^2}+\frac{\otra_2}{2 q_2^2} +\beta_0 (
q_1^2+q_2^2  )    .
\label{ai}
\ee
On the other hand, analogues of the Kepler--Coulomb (KC) potential  can be
obtained by considering any $\pot$ such that $\lim_{z\to 0} \pot =
-\cteb/\sqrt{\jm}$ ($\cteb$ is another real constant):
\be
 {\cal H}^{\rm KC} = \frac 12 \left( p_1^2 + p_2^2 \right)+
\frac{\otra_1}{2 q_1^2}+\frac{\otra_2}{2 q_2^2} -\frac{\cteb}{\sqrt{q_1^2+q_2^2 
}}.
\label{aai}
\ee

 In Section 3 we shall propose the following
{\em integrable} SW and KC systems, ${\cal H}^{\rm \SW}_z$ and ${\cal H}^{\rm
\kc}_z$,  on  the spaces of   non-constant curvature previously defined through
${\cal H}_z^{\rm I}$ (\ref{freei}):
\bea
&&{\cal H}^{\rm \SW}_z=\frac 12 \jp +
 {\beta_0} \,\frac{\sinh{z \jm}}{z} 
\label{ahad}\\
&&{\cal H}^{\rm \kc}_z=\frac 12 \jp -
 \cteb \,\sqrt{  \frac{2z}{{\rm e}^{2 z \jm}-1}} \,{\rm e}^{2 z \jm} .
\label{ahat}
\eea

The SW potential on the spaces of constant curvature  
  with free motion given by ${\cal H}_z^{\rm S}$ (\ref{frees}) will be introduced
by means of the following choice for the
 Hamiltonian
\be
 {\cal H}^{\rm \Stc}_z=\frac 12 \jp {\rm e}^{ z \jm}+
 {\beta_0} \,\frac{\sinh{z \jm}}{z} \,{\rm e}^{ z \jm}\equiv {\cal H}^{\rm
\SW}_z   {\rm e}^{ z \jm} .
\label{bh} 
\ee
We already know~\cite{Deform,CRMAngel} that this gives rise, 
  under (\ref{be}),  to a
St\"ackel-type system~\cite{Per} and  so determines a   {\em superintegrable}
deformation of    (\ref{ai}) since, besides (\ref{bf}), there exists   an
additional constant of the motion given by
\be
{\cal I}_z=\frac {\sinh z
q_1^2}{2 z q_1^2} \, {\rm e}^{z q_1^2}  p_1^2 +\frac{z \otra_1}{2\sinh z
q_1^2} \, {\rm e}^{z q_1^2}+\frac{\beta_0}{2z}\,{\rm e}^{2 z q_1^2} .
\label{bjj}
\ee
Note that this extra integral is {\it not} obtained from the coalgebra
symmetry of the Hamiltonian.\footnote{In order to construct the
superintegrable  KC system on   spaces of constant curvature, at least one of the
$\otra_1,\otra_2$ parameters must vanishes and both the specific additional
`Laplace-Runge-Lenz' integral  and the appropriate
$\pot$ compatible with it should be previously obtained.} 
Section 4 is fully devoted to the study of ${\cal H}^{\rm
\Stc}_z$,  which is shown to provide  a superintegrable system containing
a (curved) harmonic oscillator together with two more potential terms
(either oscillators or centrifugal barriers) on the usual sphere,
hyperbolic and  (anti-)de Sitter spaces. The explicit potentials are  analysed in
detail for each particular space. We stress that we recover known results on 
the Riemannian spaces~\cite{groscheS2S3,KalninsH2,RS,PogosClass1,PogosClass2} but
also we obtain new ones on  the relativistic spacetimes. Finally, some remarks and
comments close the paper.

\sect{Deformed geodesic motion}

\subsect{Integrable geodesic motion on spaces of non-constant curvature}

To start with, let us consider the metric (\ref{cc})
defined by the free Hamiltonian (\ref{freei}). Let us call
$z=\la_1^2$ and introduce another  parameter $\la_2\ne 0$,
such that $\la_1$ and $\la_2$ can be either real or
pure imaginary numbers.  Then under the change  of coordinates
$(q_1,q_2)\to (\rr,\te)$ defined by~\cite{plb} 
\be
\cosh(\la_1 \rr)=\exp\left\{z(q_1^2+q_2^2)\right\}\equiv{\rm e}^{z\jm}\qquad
 \sin^2(\la_2 \te)=\frac{\exp\left\{2z q_1^2
\right\}-1}{\exp\left\{2z(q_1^2+q_2^2)\right\}-1}
\label{ce}
\ee
the metric (\ref{cc}) and curvature (\ref{cd}) are written as
\be
\d s^2=\frac {1}{\cosh(\la_1 \rr)}
\left( \d \rr^2  +\la_2^2\,\frac{\sinh^2(\la_1 \rr)}{\la_1^2} \, \d \te^2  \right)  
\label{cg}
\ee
\be
K(\rr)=-\frac 12 \la_1^2 \,\frac{\sinh^2(\la_1 \rr)}{\cosh(\la_1 \rr)} .
\label{cj}
\ee
The   product $\cosh(\la_1
\rr)\d s^2$ coincides with the metric of the 2D Cayley--Klein (CK)
spaces \cite{Yaglom,Trigo,Conf}, all  of them with constant curvature 
$\kappa_1\equiv -z$, provided that $(\rr,\te)$ are
identified with   geodesic polar coordinates.  Hence $\la_1$, $\la_2$ play the
role of (graded) contraction parameters, determining the  curvature and the 
signature of the metric, respectively.

Consequently, the metric (\ref{cg}) can be interpreted   as a deformation
of the CK metric through the factor 
$1/\cosh(\la_1 \rr)={\rm e}^{-z\jm}$, which is responsible for the transition from 
the constant curvature   to the  non-constant one, or alternatively
as a deformation of the {\em flat} Euclidean
($\la_2$ real) or Minkowskian ($\la_2$ imaginary) spaces, which are
recovered under the classical limit $z\to 0$.
 The expressions (\ref{cg}) and   (\ref{cj}) are
explicitly written for  each  particular `deformed' space  in  table
\ref{table1}; more details   can be found in~\cite{plb}.

From (\ref{cg}) we   compute  the Christoffel  
$\Gamma^i_{jk}$,  Riemann  
$R^i_{jkl}$ and   Ricci $R_{ii}$ tensors~\cite{Doub}; their non-zero components turn
out to be
\be
\begin{array}{l}
\displaystyle{
 \Gamma^\rr_{\rr\rr}=-\frac 12\la_1\tanh(\la_1 \rr)\qquad \Gamma^\te_{\te
\rr}=\la_1\,\frac{1+\cosh^2(\la_1 \rr)}{ \sinh(2\la_1 \rr)} }\\
\displaystyle{\Gamma^\rr_{\te\te}=-\frac {\la_2^2}{2\la_1}\ \tanh(\la_1
\rr)\left(1+\cosh^2(\la_1 \rr)
\right)  }
\end{array} 
\label{ch}
\ee 
\be
R^\rr_{\te  \rr \te} =R_{\te  \te} =-\frac 12 \la_2^2\sinh^2(\la_1 \rr)\tanh^2(\la_1 \rr)
\qquad
R^\te_{\rr\te \rr}=R_{\rr\rr}=-\frac 12 \la_1^2 \tanh^2(\la_1 \rr) .
\label{ci}
\ee
The sectional and scalar curvatures are (\ref{cj}) and  $2K(\rr)$, respectively.  
The connection    (\ref{ch}) allows us to write the   geodesic
equations 
\be
\frac{\d^2 q_i}{\d s^2}+\Gamma_{jk}^i \,\frac{\d q_j}{\d s}\,\frac{\d q_k}{\d s}=0
\ee
which, in our case, give rise to the following    equations  for
$(\rr(s),\te(s))$ where $s$ is the canonical parameter of the metric (\ref{cg}):
\bea
&&\!\!\!\!\!\!\!\!\!\!\!\!\!\!\!\!\!\!\!\!\!\!\!\!\!\!\!\!\!\!
\frac{\d^2 \rr}{\d s^2}-\frac 12 \la_1\tanh(\la_1 \rr) \left( \frac{\d \rr}{\d
s}\right)^2 -\frac{\la_2^2}{2\la_1}\tanh(\la_1 \rr)  \left(  1+\cosh^2(\la_1 \rr)\right) 
\left( \frac{\d
\te}{\d s}\right)^2=0
\label{ck}\\
&&\!\!\!\!\!\!\!\!\!\!\!\!\!\!\!\!\!\!\!\!\!\!\!\!\!\!\!\!\!\!
\frac{\d^2 \te}{\d s^2}+2\la_1\left(  \frac{ 1+\cosh^2(\la_1 \rr)}{\sinh(2\la_1
\rr)}\right)  
 \frac{\d  \rr}{\d s} \, \frac{\d \te}{\d s} =0 .
\label{cl}
\eea 
The latter equation  has a first integral given by
\be
\frac{\d  }{\d s} \left( \frac{\sinh^2(\la_1 \rr)}{\la_1^2\cosh(\la_1 \rr)}\, \frac{\d
\te}{\d s}\right)=0\quad 
\Longrightarrow\quad
\frac{\sinh^2(\la_1 \rr)}{\la_1^2\cosh(\la_1 \rr)}\, \frac{\d \te}{\d
s}=-\cte
\label{cm}
\ee
where $\cte$ is a constant. By substituting this result in the metric
(\ref{cg}) we find the following velocities:
\be
\begin{array}{l}
\displaystyle{
\left(\frac{\d \rr}{\d s} \right)^2=\cosh(\la_1 \rr)-\frac{\la_1^2\la_2^2\cte^2}
{\tanh^2(\la_1 \rr)} }\qquad 
\displaystyle{
  \frac{\d \te}{\d s}=-\cte\,\frac{\la_1^2\cosh(\la_1 \rr)}{\sinh^2(\la_1 \rr)}}  
\end{array} 
\label{cn}
\ee 
and  the equation (\ref{ck}) is fulfilled.  Thus the geodesic curve
$\rr=\rr(\te)$ is obtained by eliminating the parameter $s$ and then
integrating the equation
\be
\left(\frac{\d \rr}{\d \te} \right)^2=
\frac{\sinh^2(\la_1 \rr)}{ \la_1^2 \cte^2 }
\left(\frac{\sinh^2(\la_1 \rr)}{\la_1^2\cosh(\la_1
\rr)}- \la_2^2\cte^2    \right)  
\label{co}
\ee
which finally leads to an involved  solution  that  depends on elliptic functions.

Notice that all the expressions (\ref{ck})--(\ref{co}) are well defined under
the limit
$\la_1\to 0$. In this   non-deformed/flat case we recover the known
Euclidean and Minkowskian geodesics. Explicitly, from (\ref{cn}) and by
taking the limit
$\la_1\to 0$, we find that
\be
\rr^2=s^2+\la_2^2\cte^2\qquad \frac{\tan\left(\la_2(\te+\te_0)
\right)}{\la_2}=\frac{\cte}{s}
\label{cp}
\ee
where $\te_0$ is the second integration constant. Either from (\ref{cp}) or from
(\ref{co}) with $\la_1= 0$, we obtain the curve
\be
\frac{\cte}{\rr}= \frac{\sin(\la_2(\te+\te_0))}{\la_2}  .
\label{cq}
\ee

\begin{table}[t]
{\footnotesize
 \noindent
\caption{{Metric and sectional curvature of the four spaces of
non-constant curvature for $\la_1,\la_2\in\{1,{\rm i}\}$ with deformed
$sl(2)$-coalgebra symmetry. The contraction $z=\la_1^2=0$ leads to the
flat Euclidean and Minkowskian spaces.}}
\label{table1}
\medskip
\noindent\hfill
$$
\begin{array}{ll}
\hline
\\[-6pt]
{\mbox {2D deformed Riemannian spaces}}&\quad{\mbox  {$(1+1)$D
deformed relativistic spacetimes}}\\[4pt] 
\hline
\\[-6pt]
\mbox {$\bullet$ Deformed sphere ${\bf S}^2_z$}&\quad\mbox {$\bullet$ Deformed anti-de Sitter
spacetime ${\bf AdS}^{1+1}_z$}\\[4pt] z=-1;\ (\la_1,\la_2)=({\rm i},1)&\quad
z=-1;\ (\la_1,\la_2)=({\rm i},{\rm i})\\[4pt]
 \displaystyle{\d s^2 =\frac{1}{\cos \rr}\left( \d \rr^2+\sin^2 \rr\,\d\te^2
\right)} &\quad
 \displaystyle{\d s^2 =\frac{1}{\cos \rr}\left( \d \rr^2-\sin^2 \rr\,\d\te^2
\right)} \\[8pt]
 \displaystyle{K =-\frac{\sin^2 \rr}{2\cos \rr} } &\quad
 \displaystyle{K =-\frac{\sin^2 \rr}{2\cos \rr} } \\[12pt]
\mbox {$\bullet$ Euclidean space  ${\bf E}^2$}&\quad\mbox {$\bullet$ Minkowskian spacetime ${\bf
M}^{1+1}$}\\[4pt]
z=0;\ (\la_1,\la_2)=(0,1)&\quad
z=0;\ (\la_1,\la_2)=(0,{\rm i})\\[4pt]
 \displaystyle{\d s^2 =  \d \rr^2+ \rr^2\d\te^2
 } &\quad
 \displaystyle{\d s^2 =  \d \rr^2- \rr^2\d\te^2} \\[2pt]
 \displaystyle{K =0 } &\quad
 \displaystyle{K =0} \\[6pt]
\mbox {$\bullet$ Deformed hyperbolic space ${\bf H}_z^2$}&\quad\mbox {$\bullet$ Deformed de Sitter
spacetime ${\bf dS}^{1+1}_z$}\\[4pt]
z=1;\ (\la_1,\la_2)=(1,1)&\quad
z=1;\ (\la_1,\la_2)=(1,{\rm i})\\[4pt]
 \displaystyle{\d s^2 =\frac{1}{\cosh \rr}\left( \d \rr^2+\sinh^2 \rr\,\d\te^2
\right)} &\quad
 \displaystyle{\d s^2 =\frac{1}{\cosh \rr}\left( \d \rr^2-\sinh^2 \rr\,\d\te^2
\right)} \\[8pt]
\displaystyle{K =-\frac{\sinh^2 \rr}{2\cosh \rr} } &\quad
 \displaystyle{K =-\frac{\sinh^2 \rr}{2\cosh \rr} } \\[8pt]
\hline
\end{array}
$$
\hfill}
\end{table}

\subsect{Superintegrable geodesic motion on spaces of constant curvature}

Le us now consider now the   metric (\ref{ec}) that underlies the  free
superintegrable Hamiltonian (\ref{frees}).  Under the change of
coordinates (\ref{ce}) we find that
\be
\d s^2=\frac {1}{\cosh^2(\la_1 \rr)}
\left( \d \rr^2  +\la_2^2\,\frac{\sinh^2(\la_1 \rr)}{\la_1^2} \, \d \te^2  \right)  .
\label{ee}
\ee
  By introducing a new radial coordinate $r$ defined by~\cite{plb}
\be
r=\int_0^{\rr}\frac{\d x}{\cosh(\la_1 x)}  
\label{ef}
\ee
  we  obtain  exactly  the CK metric written in  geodesic polar coordinates $(r,\te)$ provided
that  now $ \kappa_1\equiv z=\la_1^2 $~\cite{Conf}:
\be
\d s^2= 
 \d r^2  +\la_2^2\,\frac{\sin^2(\la_1 r)}{\la_1^2} \, \d \te^2  . 
\label{eh}
\ee

The non-zero connection and   curvature tensors  are given by
\be
 \Gamma^r_{\te\te}=-\la_2^2 \,\frac {\sin(\la_1 r)}{\la_1}\, {\cos(\la_1 r)}\qquad
 \Gamma^\te_{\te r}= \frac{ \la_1 }{ \tan( \la_1 r)} 
\label{ei}
\ee 
\be
R^r_{\te  r \te} =R_{\te  \te} =  \la_2^2\sin^2(\la_1 r) 
\qquad
R^\te_{r\te r}=R_{rr }=\la_1^2\equiv z  
\label{ej}
\ee
so that the sectional curvature is  just $K=z$.  By applying the very same
procedure described in Section 2.1 we deduce the 
  generic geodesic  curve $r=r(\te)$:
\be
\frac{\la_1 \cte}{\tan(\la_1 r)}= \frac{\sin(\la_2(\te+\te_0))}{\la_2}  
\label{ek}
\ee
where $\cte$ and $\te_0$ are integration constants. In the flat case
$z\to 0$ the coordinate  $\rr\to r$, so that the curve (\ref{ek}) 
 coincides with (\ref{cq}), as it should be.

We specialize all this information for each space in table \ref{table3}. 
When comparing  tables \ref{table1} and \ref{table3}, notice that the sign of $z$ for
a given `deformed' space  of non-constant curvature is the opposite to
the corresponding one    of  constant curvature (this is a consequence of
the definition (\ref{ef})).

\begin{table}[t]
{\footnotesize
 \noindent
\caption{{Metric, connection, geodesics and sectional curvature of the six spaces of  constant
curvature for $\la_1\in\{1,0,{\rm i}\}$ and  $\la_2\in\{1,{\rm i}\}$ with deformed $sl(2)$-coalgebra
symmetry.}}
\label{table3}
\medskip
\noindent\hfill
$$
\begin{array}{ll}
\hline
\\[-6pt]
{\mbox {2D Riemannian spaces}}&\quad{\mbox  {$(1+1)$D relativistic
spacetimes}}\\[4pt] 
\hline
\\[-6pt]
\mbox {$\bullet$ Sphere ${\bf S}^2$: $(\la_1,\la_2)=(1,1)$}&\quad\mbox {$\bullet$ Anti-de Sitter
spacetime
${\bf AdS}^{1+1}$: $(\la_1,\la_2)=(1,{\rm i})$}\\[4pt] 
\displaystyle{\d s^2 =   \d  r^2+\sin^2  r\,\d\te^2} &\quad
 \displaystyle{\d s^2 =  \d  r^2-\sin^2  r\,\d\te^2} \\[2pt]
 \displaystyle{  \Gamma^r_{\te\te}=- \sin r \cos  r \quad
 \Gamma^\te_{\te r}= 1/\tan  r   } &\quad
 \displaystyle{ \Gamma^r_{\te\te}=  \sin r \cos  r \quad
 \Gamma^\te_{\te r}= 1/\tan  r  } \\[2pt]
\displaystyle{  \cte/ \tan  r =  \sin(\te+\te_0)   } &\quad
 \displaystyle{ \cte/ \tan  r =  \sinh(\te+\te_0) } \\[2pt]
 \displaystyle{K =1 } &\quad
 \displaystyle{K =1} \\[6pt]
\mbox {$\bullet$ Euclidean space  ${\bf E}^2$: $(\la_1,\la_2)=(0,1)$}&\quad\mbox
{$\bullet$ Minkowskian spacetime ${\bf M}^{1+1}$: $(\la_1,\la_2)=(0,{\rm i})$}\\[4pt]
 \displaystyle{\d s^2 =  \d  r^2+  r^2\d\te^2
 } &\quad
 \displaystyle{\d s^2 =  \d  r^2-  r^2\d\te^2} \\[2pt]
\displaystyle{  \Gamma^r_{\te\te}=- r \quad
 \Gamma^\te_{\te r}= 1/r   } &\quad
 \displaystyle{ \Gamma^r_{\te\te}=  r \quad
 \Gamma^\te_{\te r}= 1/r  } \\[2pt]
\displaystyle{  \cte/  r =  \sin(\te+\te_0)   } &\quad
 \displaystyle{ \cte/ r =  \sinh(\te+\te_0) } \\[2pt]
 \displaystyle{K =0 } &\quad
 \displaystyle{K =0} \\[6pt]
\mbox {$\bullet$ Hyperbolic space ${\bf H}^2$: $(\la_1,\la_2)=({\rm i},1)$}&\quad\mbox {$\bullet$ De
Sitter spacetime ${\bf dS}^{1+1}$: $(\la_1,\la_2)=({\rm i},{\rm i})$}\\[4pt]
 \displaystyle{\d s^2 =  \d  r^2+\sinh^2  r\,\d\te^2
 } &\quad
 \displaystyle{\d s^2 =  \d  r^2-\sinh^2  r\,\d\te^2
 } \\[2pt]
\displaystyle{  \Gamma^r_{\te\te}=- \sinh r \cosh  r \quad
 \Gamma^\te_{\te r}= 1/\tanh  r   } &\quad
 \displaystyle{ \Gamma^r_{\te\te}=  \sinh r \cosh  r \quad
 \Gamma^\te_{\te r}= 1/\tanh  r  } \\[2pt]
\displaystyle{  \cte/ \tanh  r =  \sin(\te+\te_0)   } &\quad
 \displaystyle{ \cte/ \tanh  r =  \sinh(\te+\te_0) } \\[2pt]
\displaystyle{K =-1 } &\quad
 \displaystyle{K =-1 } \\[6pt]
\hline
\end{array}
$$
\hfill}
\end{table}

\sect{Integrable potentials on spaces of non-constant curvature}

Let  $(p_\rr,p_\te)$ be the canonical momenta corresponding to the new
coordinates
$(\rr,\te)$ (\ref{ce}). The relationship between the initial  
 phase space   $(q_1,q_2;p_1,p_2)$   and  the new one
$(\rr,\te;p_\rr,p_\te)$  is found to be
\be
\begin{array}{l}
\displaystyle{
\frac{p_1}{q_1}=\frac{\la_1}{\tanh(\la_1 \rr)}\,p_\rr+\frac{\la_1^2}{\la_2
\sinh^2(\la_1 \rr)\tan(\la_2\te) }\,p_\te }\\[12pt]
\displaystyle{
\frac{p_2}{q_2}=\frac{\la_1}{\tanh(\la_1 \rr)}\,p_\rr-
\frac{\la_1^2\tan(\la_2\te) }{\la_2\tanh^2(\la_1 \rr) }\,p_\te  } .
\end{array} 
\label{dc}
\ee 
Therefore if we consider the generic integrable Hamiltonian  
\be
{\cal H}_z^{\rm I}=\frac 12 \jp +
\pot(z\jm)
\label{ahadgen}
\ee
and we perform the corresponding transformations we get
\bea
&& 
{H}^{\rm I}_z=\frac 12 \cosh(\la_1
\rr)\left(p_\rr^2 +\frac{\la_1^2}{\la_2^2\sinh^2(\la_1 \rr)} \, 
p_\te^2\right)+g(\rr)\nonumber\\ 
&&\qquad\qquad +\frac{2\la_1^2 \cosh(\la_1
\rr) }{\sinh^2(\la_1 \rr) }\left(\frac{\otra_1}{\sin^2(\la_2\te)
}+\frac{\otra_2}{\cos^2(\la_2\te) }
\right)  
\label{dd}
\eea
where ${H}^{\rm I}_z=2 {\cal H}_z^{\rm I}$ and $g(\rr)=2\,\pot(z\jm(\rho))$ is an
arbitrary smooth function.
 The corresponding constant of the motion follows from (\ref{bf}). If we  
define
 $ {C}_z(\rr,\te;p_\rr, p_\te)=4\la_2^2  {\cal C}_z(q_i,p_i)$ 
we find that
\be
{C}_z=p_\te^2+\frac{4\la_2^2
\otra_1}{\sin^2(\la_2\te)}+\frac{4\la_2^2 \otra_2}{\cos^2(\la_2\te)} 
\label{de}
\ee
which does not depend on $(\rr,p_\rr)$. Furthermore this constant  allows
us to reduce (\ref{dd}) to the 1D (radial) Hamiltonian given by

\be 
{H}^{\In}_z(\rr,p_\rr)=\frac 12 \cosh(\la_1
\rr)\, p_\rr^2  +\frac{\la_1^2 \cosh(\la_1
\rr)}{2\la_2^2\sinh^2(\la_1 \rr)} \, {C}_z+ g(\rr) .
\label{dede}
\ee

Consequently, the Hamiltonian (\ref{dd}) defines a family of integrable systems which, for any
$g(\rr)$,
  share the same  constant of the motion (\ref{de}). We specialize in
table \ref{table2} these expressions for each of the six spaces shown in
table \ref{table1}.

\begin{table}[t]
{\footnotesize
 \noindent
\caption{{Integrable Hamiltonians and their constant of the motion on   the six
spaces  given in table \ref{table1}. The same conventions are
followed.}}
\label{table2}
\medskip
\noindent\hfill
$$
\begin{array}{ll}
\hline
\\[-6pt]
\mbox {$\bullet$ Deformed sphere ${\bf S}^2_z$}&\quad\mbox {$\bullet$ Deformed anti-de Sitter
spacetime ${\bf AdS}^{1+1}_z$}\\[4pt]
 \displaystyle{
{H}^{\In}_z=\frac 12 \cos  \rr \left(p_\rr^2
+\frac{1}{ \sin^2  \rr} \,  p_\te^2\right)+g(\rr) }&\quad
 \displaystyle{
{H}^{\In}_z=\frac 12 \cos  \rr \left(p_\rr^2
-\frac{1}{ \sin^2  \rr} \,  p_\te^2\right)+g(\rr) }\\[8pt]
 \displaystyle{\qquad\quad
+\frac{2 \cos
\rr }{\sin^2  \rr}\left( \frac{\otra_1}{\sin^2\te}+\frac{\otra_2}{\cos^2 \te}  
\right)} &\quad
 \displaystyle{\qquad\quad
-\frac{2 \cos
\rr }{\sin^2  \rr}\left(\frac{\otra_1}{\sinh^2\te
}-\frac{\otra_2}{\cosh^2 \te}\right)  }\\[8pt]
 \displaystyle{\qquad\, =\frac 12 \cos  \rr \, p_\rr^2
+\frac{\cos  \rr}{2 \sin^2  \rr} \,{C}_z +g(\rr) } &\quad
 \displaystyle{\qquad\, =\frac 12 \cos  \rr \, p_\rr^2
-\frac{\cos  \rr}{2 \sin^2  \rr} \,{C}_z +g(\rr)  }\\[10pt]
 \displaystyle{ {C}_z=p_\te^2+\frac{4 
\otra_1}{\sin^2 \te }+\frac{4 \otra_2}{\cos^2 \te }   }&\quad
 \displaystyle{ {C}_z=p_\te^2+\frac{4 
\otra_1}{\sinh^2 \te }-\frac{4 \otra_2}{\cosh^2 \te }  }\\[14pt]

 \mbox {$\bullet$ Euclidean space  ${\bf E}^2$}&\quad\mbox {$\bullet$ Minkowskian spacetime ${\bf
M}^{1+1}$}\\[4pt]
\displaystyle{
{H}^{\In} =\frac 12   \left(p_\rr^2
+\frac{1}{\rr^2} \,  p_\te^2\right)+g(\rr)  }&\quad
 \displaystyle{
{H}^{\In} =\frac 12   \left(p_\rr^2
-\frac{1}{   \rr^2} \,  p_\te^2\right)+g(\rr) }\\[8pt]
\displaystyle{\qquad\quad 
+\frac{2}{ \rr^2}\left( \frac{\otra_1}{\sin^2\te}+\frac{\otra_2}{\cos^2
\te}   \right) }&\quad
 \displaystyle{ \qquad\quad
 -\frac{2   }{ \rr^2}\left(\frac{\otra_1}{\sinh^2\te
}-\frac{\otra_2}{\cosh^2 \te}\right)  }\\[8pt] 
 \displaystyle{\qquad\, =\frac 12   \, p_\rr^2
+\frac{1}{2 \rr^2} \,{C}_z +g(\rr) } &\quad
 \displaystyle{\qquad\, =\frac 12  \, p_\rr^2
-\frac{1}{2 \rr^2} \,{C}_z +g(\rr)  }\\[10pt]
 \displaystyle{ {C} =p_\te^2+\frac{4 
\otra_1}{\sin^2 \te }+\frac{4 \otra_2}{\cos^2 \te }   }&\quad
 \displaystyle{ {C} =p_\te^2+\frac{4 
\otra_1}{\sinh^2 \te }-\frac{4 \otra_2}{\cosh^2 \te }  }\\[14pt]

 \mbox {$\bullet$ Deformed hyperbolic space ${\bf H}_z^2$}&\quad\mbox {$\bullet$ Deformed de Sitter
spacetime ${\bf dS}^{1+1}_z$}\\[4pt]
\displaystyle{
{H}^{\In}_z=\frac 12 \cosh  \rr \left(p_\rr^2
+\frac{1}{ \sinh^2  \rr} \,  p_\te^2\right)+g(\rr) }&\quad
 \displaystyle{
{H}^{\In}_z=\frac 12 \cosh  \rr \left(p_\rr^2
-\frac{1}{ \sinh^2  \rr} \,  p_\te^2\right)+g(\rr) }\\[8pt]
 \displaystyle{\qquad\quad
+\frac{2 \cosh
\rr }{\sinh^2  \rr}\left( \frac{\otra_1}{\sin^2\te}+\frac{\otra_2}{\cos^2 \te}  
\right)} &\quad
 \displaystyle{\qquad\quad
-\frac{2 \cosh
\rr }{\sinh^2  \rr}\left(\frac{\otra_1}{\sinh^2\te
}-\frac{\otra_2}{\cosh^2 \te}\right)  }\\[8pt]
 \displaystyle{\qquad\, =\frac 12 \cosh  \rr \, p_\rr^2
+\frac{\cosh  \rr}{2 \sinh^2  \rr} \,{C}_z +g(\rr) } &\quad
 \displaystyle{\qquad\, =\frac 12 \cosh  \rr \, p_\rr^2
-\frac{\cosh  \rr}{2 \sinh^2  \rr} \,{C}_z +g(\rr)  }\\[10pt]
 \displaystyle{ {C}_z=p_\te^2+\frac{4 
\otra_1}{\sin^2 \te }+\frac{4 \otra_2}{\cos^2 \te }   }&\quad
 \displaystyle{ {C}_z=p_\te^2+\frac{4 
\otra_1}{\sinh^2 \te }-\frac{4 \otra_2}{\cosh^2 \te }  }\\[8pt]
\hline
\end{array}
$$
\hfill}
\end{table}

\subsect{The SW potential}

Amongst the possible choices for the  deformed Hamiltonian (\ref{ahadgen}), let
us consider the   expression (\ref{ahad}) whose non-deformed
limit is the SW system (\ref{ai}).
This choice implies that  the  potential  function $g(\rr)$ appearing in (\ref{dd})
reads\be
g(\rr)=\beta_0\cosh(\la_1 \rr)\,\frac{\tanh^2(\la_1 \rr)}{\la_1^2} .
\label{df}
\ee
This gives
\be 
 {H}^{\SW}_z=\cosh(\la_1 \rr) {H}^{\SStc}_z
\label{dg}
\ee 
where    
\bea   
&& {H}^{\SStc}_z  =\frac 12 \left(p_\rr^2
+\frac{\la_1^2}{\la_2^2\sinh^2(\la_1 \rr)} \,  p_\te^2\right)+\beta_0\,
\frac{\tanh^2(\la_1 \rr)}{\la_1^2} \nonumber\\
&&\qquad\qquad  +\frac{2\la_1^2  }{\sinh^2(\la_1 \rr)
}\left(\frac{\otra_1}{\sin^2(\la_2\te) }+\frac{\otra_2}{\cos^2(\la_2\te) }
\right) .
\label{dh}
\eea  

The physical interpretation of (\ref{dg}) is the following.

\begin{itemize}

\item[$\bullet$]  When $z=0$ $(\la_1\to 0)$, the expression  (\ref{dg})  reduce to
the
 SW Hamiltonian on   ${\bf E}^2$    and  ${\bf
M}^{1+1}$   written in polar coordinates
\be   
{H}^{\SW}\equiv {H}^{\SStc}  =\frac 12 \left(p_\rr^2
+\frac{ p_\te^2}{\la_2^2  \rr^2} \, \right)+\beta_0\,
\rr^2   +\frac{2}{\rr^2}\left(\frac{\otra_1}{\sin^2(\la_2\te)
}+\frac{\otra_2}{\cos^2(\la_2\te) }
\right)  .
\label{di}
\ee  
The
term $\beta_0 \rr^2$ is the usual harmonic oscillator potential, while those
depending on  $\otra_1$ and   $\otra_2$ are two `centrifugal barriers'.

 \item[$\bullet$] 
 When $z\ne 0$, $ {H}^{\SStc}_z$ is well-known in the literature as the
superintegrable SW system on spaces of constant curvature~\cite{
groscheS2S3,KalninsH2,RS,PogosClass1,PogosClass2}, where  
the $\beta_0$-term   corresponds to a `curved' harmonic oscillator
potential.
 In particular, if $z=-1$ we
recover the spherical oscillator or Higgs potential~\cite{Higgs,Leemon},  
$\beta_0 \tan^2\rr$, on  
${\bf S}^2$  and  ${\bf AdS}^{1+1}$, while $z=1$
leads to a  hyperbolic oscillator $\beta_0 
\tanh^2\rr$ on   ${\bf H}^2$  and  ${\bf dS}^{1+1}$. 
  Hence
$ {H}^{\SW}_z$ can  properly be regarded as an  integrable generalization
of the superintegrable ${H}^{\SStc}_z $   to spaces of non-constant
curvature. We stress that ${H}^{\SStc}_z $ can be  reproduced from the  
Hamiltonian (\ref{bh}) in an exact way and it will be analysed   in the
next section.

 \end{itemize}

\subsect{The KC potential}

If  we now consider the Hamiltonian (\ref{ahat}) 
whose non-deformed limit is the KC system (\ref{aai}), we get
\be
g(\rr)=-\ctec\cosh(\la_1 \rr)\,\frac{\la_1}{\tanh(\la_1 \rr)} 
\label{dj}
\ee
with $\ctec=2\sqrt{2}\,\cteb$. With this choice we find that
\be 
  {H}^{\rm IKC}_z=\cosh(\la_1 \rr) {H}^{\rm KC}_z
\label{dl} 
\ee
where
\bea
&& {H}^{\rm KC}_z  =\frac 12 \left(p_\rr^2
+\frac{\la_1^2}{\la_2^2\sinh^2(\la_1 \rr)} \,  p_\te^2\right)-
\ctec\,
\frac{\la_1}{\tanh(\la_1 \rr)}\label{ddll} \\
&&\qquad\qquad  +\frac{2\la_1^2  }{\sinh^2(\la_1 \rr)
}\left(\frac{\otra_1}{\sin^2(\la_2\te) }+\frac{\otra_2}{\cos^2(\la_2\te) }
\nonumber
\right) 
\eea
and the integrable (but not superintegrable) Hamiltonian ${H}^{\rm KC}_z$
contains a KC potential in polar coordinates through the $\ctec$-term.
Explicitly,

\begin{itemize}

\item[$\bullet$] When $z=0$ we obtain   the reduction
\be   
{H}^{\rm IKC}\equiv {H}^{\rm KC}  =\frac 12 \left(p_\rr^2
+\frac{ p_\te^2}{\la_2^2  \rr^2} \, \right)-\frac{\ctec}
{\rr}  +\frac{2}{\rr^2}\left(\frac{\otra_1}{\sin^2(\la_2\te)
}+\frac{\otra_2}{\cos^2(\la_2\te) }
\right)  
\label{ddi}
\ee  
which  defines an integrable
system formed by a composition of  the KC potential, $-\ctec/\rr$, with two
centrifugal barriers on the flat spaces  ${\bf E}^2$ and  ${\bf
M}^{1+1}$. This  Hamiltonian is superintegrable   whenever at
least one of the parameters $\otra_1$, $\otra_2$ is taken equal
to zero (see, e.g.,~\cite{RS,Evans}).

\item[$\bullet$]  On the contrary, if   $z\ne 0$ then
${H}^{\rm IKC}_z\ne {H}^{\rm KC}_z$ and the latter contains the   KC
potential  on spaces of constant
curvature~\cite{RS,PogosClass1,PogosClass2,Schrodingerdual,Schrodingerdualb,Schrodingerdualc}
in  its   spherical version,
$-\ctec/\tan\rr$,  on ${\bf S}^2$~\cite{Schrodinger} and   ${\bf AdS}^{1+1}$
$(z=-1)$, as well as in
 its hyperbolic one, $-\ctec/\tanh\rr$, on ${\bf H}^2$ and   ${\bf
dS}^{1+1}$  $(z=1)$. Therefore we can conclude that    
 $ {H}^{\rm IKC}_z$ (\ref{dl}) is an appropriate generalization of the   KC
potential to spaces of non-constant curvature, possibly supplemented by two more
potential terms.

\end{itemize}


\sect{The superintegrable SW potential on spaces with\\ constant curvature}

Now we can reproduce the same study in the case of the superintegrable Hamiltonian
(\ref{bh}) defined on the spaces of constant curvature of Section 2.2.

The relation 
between  the initial phase space  
$(q_i,p_i)$   and  the   canonical geodesic
polar coordinates $(r,\te)$ and momenta $(p_r,p_\te)$   turns out to be
\be
\begin{array}{l}
\displaystyle{
\frac{p_1}{q_1}=\frac{\la_1}{\tan(\la_1 r)}\,p_r+\frac{\la_1^2}{\la_2
\tan^2(\la_1 r)\tan(\la_2\te) }\,p_\te }\\[12pt]
\displaystyle{
\frac{p_2}{q_2}=\frac{\la_1}{\tan(\la_1 r)}\,p_r-
\frac{\la_1^2\tan(\la_2\te) }{\la_2\sin^2(\la_1 r) }\,p_\te  } .
\end{array} 
\label{fj}
\ee 
By considering the realization (\ref{be}), the   change of
coordinates defined through the composition of (\ref{ce}) and
(\ref{ef}) together with  the above relations, it can be shown
that the  Hamiltonian
${\cal H}^{\rm \Stc}_z$    (\ref{bh}) and its constants of the
motion  
${\cal C}_z$ (\ref{bf}) and  ${\cal I}_z$ (\ref{bjj}) are expressed in the phase space
$(r,\te;p_r,p_\te)$  as 
\bea
&& 
{H}^{\rm SW}_z=\frac 12 \left(p_r^2 +\frac{\la_1^2}{\la_2^2\sin^2(\la_1 r)}
\,  p_\te^2\right)+ \beta_0 \,\frac{\tan^2(\la_1 r)}{\la_1^2} \nonumber\\ 
&&\qquad\qquad + \frac{  \la_1^2  }{\sin^2(\la_1 r) }  \left(
 \frac{\beta_1}{\cos^2(\la_2\te)} + \frac{\beta_2}{\sin^2(\la_2\te)}   
\right)  
\label{fl}\\
&& {C}_z=p_\te^2+ \frac{2  \la_2^2 \beta_1}{ \cos^2(\la_2\te) }
 +\frac{ 2 \la_2^2 \beta_2}{\sin^2(\la_2\te)}
\label{fm}\\
&&{I}_z=\left(\la_2\sin(\la_2\te) p_r+\frac{\la_1\cos(\la_2\te)}{\tan(\la_1 r)}\,p_\te
\right)^2\nonumber\\
&&\qquad\qquad +2\beta_0\la_2^2\,\frac{\tan^2(\la_1 r)}{\la_1^2 }\,\sin^2(\la_2\te)
+\frac{2\beta_2\la_1^2\la_2^2}{ \tan^2(\la_1 r)\sin^2(\la_2\te)}
\label{fn}
\eea
where we have rescaled these quantities in the following way:
\be
\begin{array}{l}
\displaystyle{ {H}^{\SStc}_z=2{\cal H}^{\Stc}_z\quad\ 
{C}_z=4\la_2^2 {\cal C}_z 
\quad\
{I}_z=4\la_2^2\left({\cal I}_z-\frac{\beta_0}{2\la_1^2} -\la_1^2 \otra_1\right)
}\\
 \beta_1=2\otra_2\qquad \beta_2=2\otra_1.
\end{array}
\label{fk}
\ee
The constant of the motion ${C}_z$ allows us   to  reduce 
 ${H}^{\SStc}_z$ to a 1D radial  system given by
\be 
{H}^{\SStc}_z=\frac 12 \, p_r^2 +\frac{\la_1^2}{2\la_2^2\sin^2(\la_1 r)} \, 
 {C}_z + \beta_0 \,\frac{\tan^2(\la_1 r)}{\la_1^2}   .
\label{ffnn}
\ee

These results are displayed in table
\ref{table4} for each space of constant curvature.
The superintegrable Hamiltonians written on the first column of table \ref{table4} are
constructed   on the three classical {\em Riemannian spaces} ($\la_2=1$) and they are already
known~\cite{groscheS2S3,KalninsH2,RS,PogosClass1,PogosClass2}. We
recall that all these results were obtained by applying
 different procedures that are not related to coalgebra symmetry. In
geodesic polar coordinates these systems were also constructed
in~\cite{VulpiLett,CRMVulpi} through a Lie group approach, where the
constants of the motion there denoted $I_{12}$, $I_{02}$ are related
to    
 (\ref{fm}) and (\ref{fn}) by  $I_{12}= {C}_z - 2\la_2^2(\beta_1+\beta_2) $, 
$I_{02}={I}_z$. The Hamiltonian ${H}^{\SStc}_z$   on
${\bf S}^2$ has been interpreted
in~\cite{ran,ran1,ran2,ranran}  as a superposition of three spherical oscillators; the
corresponding geometrical interpretation on ${\bf H}^2$ can be found in~\cite{CRMVulpi}.

We stress that the second column of table \ref{table4} provides the
coalgebraic generalization of the SW potential to the three `classical' {\em
relativistic spacetimes} ($\la_2={\rm i}$)  which,  to our knowledge, was
still lacking. In the following we shall describe the system
${H}^{\SStc}_z$ on the six spaces, that includes a full description of the
new cases (on
${\bf AdS}^{1+1}$, ${\bf M}^{1+1}$ and 
${\bf dS}^{1+1}$), as well as of the known potentials on  ${\bf
S}^2$, ${\bf E}^2$  and  ${\bf H}^2$. We point out that  
  when the six   spaces are considered altogether the 
interpretation becomes more comprehensible and transparent.

\subsect{Geometrical interpretation of the SW potential}

Let us firstly recall which is the  (physical) geometrical role of the geodesic polar
coordinates $(r,\te)$ on the  spaces of constant curvature of table \ref{table3}. These
can be embedded in a 3D linear ambient  space with coordinates $(x_0,x_1,x_2)$ given by~\cite{Conf} 
\be
x_0=\cos(\la_1 r)\qquad x_1=\frac{\sin(\la_1 r)}{\la_1}\, \cos(\la_2\te)\qquad 
x_2=\frac{\sin(\la_1 r)}{\la_1}\, \frac{\sin(\la_2\te)}{\la_2}
\label{ga}
\ee
which fulfil  the
constraint $x_0^2+\la_1^2(x_1^2+\la_2^2 x_2^2)=1$; the origin has ambient
coordinates $O=(1,0,0)$. Consider a (time-like) geodesic $l_1$, another 
(space-like) geodesic $l_2$ orthogonal at $O$ and a generic point $Q$ with
coordinates  (\ref{ga}).  Then  $r$ is
the (time-like) distance along the geodesic $l$ that joins $Q$ with $O$, while $\te$ is
the (rapidity) angle  of $l$ with respect to $l_1$ as shown in figure 1. 
In the  Riemannian cases with $\la_2=1$  the coordinates 
$(r,\te)$  parametrize the  complete space, while in the relativistic cases
 with $\la_2={\rm i}$, it is verified that  $|x_2|\le |x_1|$, so that $(r,\te)$ only
cover the time-like lines limited by  the isotropic lines  $x_2=\pm x_1$  
on which $\te\to \infty$.

Next, let $Q_i$ ($i=1,2$) be the intersection point 
of $l_i$ with    its  orthogonal geodesic    through $Q$.
Then if $x$ denotes the (time-like) distance $Q\YY_2$ and $y$   the (space-like)
distance  $Q\YY_1$, it can be shown that~\cite{Conf}
\be
  x_1=\frac{\sin(\la_1 r)}{\la_1}\, \cos(\la_2\te)\equiv \frac{\sin(\la_1 x)}{\la_1}
\qquad 
x_2=\frac{\sin(\la_1 r)}{\la_1}\, \frac{\sin(\la_2\te)}{\la_2}\equiv
\frac{\sin(\la_1\la_2 y)}{\la_1\la_2} .
\label{gb}
\ee
Hence  the   potential of ${H}^{\SStc}_z$ (\ref{fl}) can be
rewritten 
 as
\be
{U}^{\SStc}_z=\beta_0\, \frac{\tan^2(\la_1 r)}{\la_1^2}  + \frac{  \la_1^2 
\beta_1}{\sin^2(\la_1 x) }  
  + \frac{ \la_1^2  \beta_2}{\sin^2(\la_1\la_2 y)}   
\label{gc}
 \ee
which conveys   a common interpretation  on the six
spaces:

\noindent
$\bullet$ The $\beta_0$-term   is a {\it central} harmonic oscillator, that is, with center
at the origin $O$.

\noindent
$\bullet$  Both $\beta_i$-terms $(i=1,2)$ are `centrifugal  barriers'.

Alternatively, the $\beta_i$-terms may adopt a different  interpretation   in each
particular space, depicted   in figure 1, that  we proceed to study. 
All the trigonometric relations on  these spaces   that are  necessary in our
description can be found in~\cite{Trigo}.


\subsubsect{Sphere ${\bf S}^2$}

  Let 
$O_i$ $(i=1,2)$ be the intersection points  of the geodesics $l_i$ with the equator of
the sphere  (at a distance $\frac \pi 2$ from $O$) and $r_i$ the  distances along the
geodesics joining
$Q$ and $O_i$. By applying the cosine theorem  for the two triangles 
$OQO_i$ we find that
\be
\begin{array}{l}
\cos r_1= \cos \frac{\pi}2 \cos r+\sin \frac{\pi}2\sin r\cos\te\\
\cos r_2= \cos
\frac{\pi}2 \cos r+\sin \frac{\pi}2\sin r\cos(\frac{\pi}2-\te)
 \end{array} 
\label{gd}
\ee
that is
\be
 \cos r_1=  \sin r\cos\te\equiv \sin x\qquad 
\cos r_2=  \sin r\sin \te \equiv \sin y.
\label{ge}
\ee
The same result follows by 
noticing that $r_1+x=r_2+y=\frac{\pi}2$ and that  $\{O_1,Q,\YY_2\}$ and 
 $\{O_2,Q,\YY_1\}$ lie  on a same geodesic orthogonal to $l_2$ and $l_1$, respectively.

Therefore the     potential (\ref{gc})  on ${\bf S}^2$ can be expressed in two ways
\bea
&&{U}^{\SStc}_z=\beta_0\tan^2 r+\frac{\beta_1}{\sin^2
x}+\frac{\beta_2}{\sin^2 y} \label{gf}\\
&&\qquad\  =   \beta_0\tan^2 r+\beta_1\tan^2 r_1+\beta_2\tan^2
r_2+\beta_1+\beta_2 
\label{gg}
\eea
which  show a superposition of the central spherical oscillator  either with
two spherical centrifugal barriers, or with two  spherical oscillators with centers 
 placed at   $O_i$~\cite{VulpiLett,CRMVulpi,ran,ran1,ran2,ranran}.


\subsubsect{Hyperbolic space ${\bf H}^2$}

In this case,  the analogous points to the previous `centers'
$O_i$   would be beyond the `actual' hyperbolic space and so
placed  in the `ideal' one~\cite{ran,ran1,ran2} (in
the exterior region of ${\bf H}^2$). Thus we only write   the   potential in the
form (\ref{gc}):
\be
{U}^{\SStc}_z=\beta_0\tanh^2 r+\frac{\beta_1}{\sinh^2
x}+\frac{\beta_2}{\sinh^2 y} 
\label{gh}
\ee
which corresponds to a superposition of a hyperbolic oscillator centered at the origin
with two hyperbolic `centrifugal barriers'~\cite{CRMVulpi}.


\subsubsect{Euclidean space ${\bf E}^2$}

The
limit $z\to 0$ of the potential ${U}^{\SStc}_z$  on   ${\bf S}^2$ and ${\bf H}^2$ 
can properly be performed on both expressions (\ref{gf}) and (\ref{gh}) leading to 
the usual  harmonic oscillator $\beta_0 r^2=\beta_0(x^2+y^2)$  and centrifugal potentials
$\beta_1  / x^2$  and $\beta_2/  y^2$, in such a manner that
$(x,y)$ are Cartesian coordinates on ${\bf E}^2$. Thus the proper SW
system is recovered~\cite{Fris,Evansa,Evansb,Groschea}.
Nevertheless, this contraction
cannot directly be applied to the   potential   on ${\bf S}^2$ written in terms of $r_i$
(\ref{gg}); in fact,   when $z\to 0$ the points $O_i\to \infty$.


\subsubsect{Anti-de Sitter spacetime ${\bf AdS}^{1+1}$}

We consider the intersection point $O_1$ between the time-like geodesic
$l_1$ and the  axis $x_1$ of the ambient space,  which is at a time-like distance $\frac
\pi 2$ from the origin
$O$.  If $r_1$  denotes  the
time-like distance  $QO_1$, by applying the cosine
 theorem  on the triangle
$OQO_1$, we find that 
\be
\begin{array}{l}
\cos r_1=  \cos \frac \pi 2\cos r + \sin \frac \pi 2 \sin r\cosh\te 
\quad \Longrightarrow\quad 
\cos r_1=    \sin r\cosh\te \equiv \sin x .
\end{array} 
\label{gi}
\ee
that is, $r_1+x=\frac \pi 2$.   Therefore the   potential becomes
\bea
&&{U}^{\SStc}_z=\beta_0\tan^2 r+\frac{\beta_1}{\sin^2
x}-\frac{\beta_2}{\sinh^2 y} \label{gj}\\
&&\qquad\  =  \beta_0\tan^2 r+\beta_1\tan^2 r_1-\frac{\beta_2}{\sinh^2 y}
+\beta_1   .
\label{gk}
\eea
The former expression exhibits  a superposition of  a time-like (spherical) oscillator
  centered at $O$  with a time-like (spherical) centrifugal potential and a
space-like (hyperbolic) one. Under the latter  form, the  
time-like centrifugal term is transformed into another  
  spherical oscillator
now with center at $O_1$.


\subsubsect{De Sitter spacetime ${\bf dS}^{1+1}$}

Recall that ${\bf AdS}^{1+1}$ and  ${\bf dS}^{1+1}$ are related by means of an
  interchanging between time-like lines and space-like ones; the former are compact
(circular) on  
${\bf AdS}^{1+1}$ and non-compact (hyperbolic) on ${\bf dS}^{1+1}$, while the latter
are  non-compact on   ${\bf AdS}^{1+1}$ but  compact on ${\bf dS}^{1+1}$.
So, conversely to previous
case,   we consider the intersection point $O_2$ between the space-like geodesic
$l_2$ and the  axis $x_2$ (at a space-like distance $\frac \pi 2$ from   $O$), such that
  $r_2$ is  the space-like distance  $QO_2$. The
cosine  theorem applied to  the triangle 
$OQQ_2$ (with external angle $\te$) gives rise to
\be
\begin{array}{l}
\cos r_2=  \cos \frac \pi 2\cosh r + \sin \frac \pi 2 \sinh r\sinh\te 
\quad \Longrightarrow\quad 
\cos  r_2= \sinh r\sinh\te \equiv \sin y.
\end{array} 
\label{gl}
\ee
Note that $r_2+y=\frac{\pi}2$. Hence we again find two ways to expressed   the    
potential on ${\bf dS}^{1+1}$
\bea
&&{U}^{\SStc}_z=\beta_0\tanh^2 r+\frac{\beta_1}{\sinh^2
x}-\frac{\beta_2}{\sin^2 y} \label{gm}\\
&&\qquad\  =  \beta_0\tanh^2 r+\frac{\beta_1}{\sinh^2 x}-\beta_2\tan^2 r_2
-\beta_2     
\label{gn}
\eea
that is, a superposition of a central time-like (hyperbolic) oscillator with
a time-like (hyperbolic) centrifugal barrier, and  either with  another space-like
(spherical) centrifugal barrier or with a   space-like (spherical)
oscillator   centered at $O_2$.


\subsubsect{Minkowskian spacetime ${\bf M}^{1+1}$}

The
limit $z\to 0$ of (\ref{gj}) and (\ref{gm})  provides the    
corresponding   SW potential on ${\bf M}^{1+1}$, which is formed by a time-like
harmonic oscillator $\beta_0 r^2=\beta_0(x^2-y^2)$, one time-like  centrifugal
barrier $\beta_1  / x^2$ together with another  space-like one    $\beta_2  / y^2$.
The coordinates $(x,y)$ are the usual time and space ones. On the contrary, the
expressions (\ref{gk}) and (\ref{gn}) are not well defined when 
$z\to 0$ since   the points $O_1$ and $O_2$ go to infinity.

\sect{Concluding remarks}

Throughout this paper we have constructed several 
Hamiltonians that belong to the quite general family ${\cal H}_z={\cal
H}_z(\jm,\jp,\jj)$ within the realization (\ref{be}). There are
certainly many other possible choices that lead to integrable systems on spaces of
non-constant curvature with deformed
$sl(2)$ coalgebra symmetry. Nevertheless,   the explicit form of such a general
Hamiltonian can be restricted by taking into account the following requirements:
 (i)~the kinetic energy is quadratic in the momenta, (ii)~the limit $z\to 0$ of the
underlying deformed spaces leads to
$\>E^2$, (iii)~dimensions of the deformation parameter are
$[z]=[J_-]^{-1}=[q_i]^{-2}$, and (iv)~the potential only depends on the
coordinates.  Under these assumptions   the most general integrable Hamiltonian  is
just (\ref{ahaa});  
  notice that the case ${\cal H}_z=\jj^2$ is transformed into a particular case of
(\ref{ahaa}) through the Casimir ${\cal C}_z$.
 Hence the (deformed) kinetic energy and
potential of the resulting Hamiltonian, ${\cal H}_z={\cal T}_z+{\cal
V}_z$, turn out to be
\bea 
&& {\cal T}_z =\frac12 \left( \frac {\sinh z
q_1^2}{z q_1^2} \, {\rm e}^{z q_2^2} p_1^2   +
\frac {\sinh z q_2^2}{z q_2^2} \, {\rm e}^{-z q_1^2} p_2^2  \right)
f\left(z(q_1^2+q_2^2)\right) 
\label{zb}\\
 &&{\cal V}_z =\left( \frac{z \otra_1}{2\sinh z
q_1^2} \,  {\rm e}^{z q_2^2} + \frac{z \otra_2}{2\sinh z
q_2^2} \,  {\rm e}^{-z q_1^2}\right) f\left(z(q_1^2+q_2^2)\right) 
+\pot\left(z(q_1^2+q_2^2)
\right)  .
\label{zc}
\eea 
We remark that the general expression for the Gaussian curvature of the 2D space associated with
the kinetic energy (\ref{zb}) can be found in~\cite{plb}. Then it can be checked that
in order to obtain a
 space of constant curvature from ${\cal T}_z$ the
choice is quite singular (e.g., $f={\rm e}^{\pm z J_-}$),
and in general one obtains very involved spaces of
non-constant curvature for which the simplest choice  is the
one we have developed in this paper with $f\equiv 1$. However
other deformed spaces underlying (\ref{zb}) and particular
potentials contained in (\ref{zc}) could be worth to be
studied. Other approaches to superintegrability on 2D spaces of variable curvature
can be found in~\cite{KKWa,KKWb}.

On the other hand, we stress that by introducing a second  parameter $\la_2$, that
determines the  signature of the metric, we have been able to obtain  integrable
potentials on relativistic spacetimes of non-constant curvature; in this context,
the deformed KC potential could be of interest in classical gravity.
Furthermore the  known superintegrable SW potential on Riemannian spaces have also
been implemented on the three classical relativistic spacetimes of
constant curvature.  Notice that we have avoided the contraction
$\la_2=0$, which is well defined on  both metrics (\ref{cg}) and (\ref{ee}), since 
this would give rise to degenerate (Newtonian) metrics,  whose dynamical
contents are not so interesting.

  Finally, we recall that the existence of an underlying 
coalgebra  symmetry for all  these two-particle
Hamiltonians ensures that they can be generalized
to $N$-dimensional spaces through the coproduct. In fact,
the corresponding  expressions  in terms of the
initial phase space
$(q_i,p_i)$ can be found in~\cite{Deform}. Nevertheless the
geometrical and physical description of the corresponding 
Hamiltonians  on $N$D curved Riemannian and relativistic
spaces  (thus including a proper study of sectional
curvatures) remains as an open problem which is currently
under investigation.

\newpage

\section*{Acknowledgements}

This work was partially supported  by the Ministerio de
Educaci\'on y Ciencia   (Spain, Project FIS2004-07913),  by the Junta de Castilla
y Le\'on   (Spain, Project  BU04/03), and by the INFN-CICyT (Italy-Spain).


\begin{table}[t]
{\footnotesize
 \noindent
\caption{{Superintegrable   SW Hamiltonians and their two constants of the motion
on   the six spaces of  constant curvature    of table \ref{table3} and with the
same conventions.}}
\label{table4}
\medskip
\noindent\hfill
$$
\begin{array}{ll}
\hline
\\[-6pt]

\mbox {$\bullet$ Sphere ${\bf S}^2$}&\quad\mbox {$\bullet$ Anti-de Sitter
spacetime
${\bf AdS}^{1+1}$}\\[4pt] 
 \displaystyle{
{H}^{\SStc}_z=\frac 12   \left(p_r^2
+\frac{1}{ \sin^2  r} \,  p_\te^2\right)+\beta_0\tan^2 r }&\quad
 \displaystyle{
{H}^{\SStc}_z=\frac 12   \left(p_r^2
-\frac{1}{ \sin^2  r} \,  p_\te^2\right)+\beta_0\tan^2 r }\\[8pt]
 \displaystyle{\qquad\quad
+\frac{1}{\sin^2  r}\left( \frac{\beta_1}{\cos^2\te}+\frac{\beta_2}{\sin^2 \te}  
\right)} &\quad
 \displaystyle{\qquad\quad
+\frac{1}{\sin^2  r}\left( \frac{\beta_1}{\cosh^2\te}-\frac{\beta_2}{\sinh^2 \te}  
\right) }\\[8pt]
 \displaystyle{\qquad\ \,= \frac 12 \,p_r^2 
+\frac{1}{2\sin^2  r}\, {C}_z+ \beta_0\tan^2 r  } &\quad
 \displaystyle{\qquad\ \,= \frac 12 \,p_r^2 
-\frac{1}{2\sin^2  r}\, {C}_z+ \beta_0\tan^2 r  }\\[10pt]
 \displaystyle{ {C}_z=p_\te^2+ \frac{2\beta_1}
{\cos^2 \te}  +\frac{2 \beta_2}{\sin^2 \te }   }&\quad
 \displaystyle{ {C}_z=p_\te^2- \frac{2\beta_1}{
\cosh^2 \te }+\frac{2 \beta_2}{\sinh^2 \te }    }  \\[10pt]
 \displaystyle{ {I}_z=\left(\sin\te \,p_r+\frac{\cos\te}{\tan r}\,p_\te \right)^2
   }&\quad
 \displaystyle{  {I}_z=\left(\sinh\te \,p_r-\frac{\cosh\te}{\tan r}\,p_\te \right)^2
}\\[8pt]
 \displaystyle{ \qquad
+2\beta_0\tan^2 r\sin^2\te +\frac{2\beta_2}{\tan^2 r\sin^2\te}   }&\quad
 \displaystyle{ \qquad
+2\beta_0\tan^2 r\sinh^2\te +\frac{2\beta_2}{\tan^2 r\sinh^2\te}   }\\[14pt]

 \mbox {$\bullet$ Euclidean space  ${\bf E}^2$}&\quad\mbox
{$\bullet$ Minkowskian spacetime ${\bf M}^{1+1}$}\\[4pt]
 \displaystyle{
{H}^{\SStc}=\frac 12   \left(p_r^2
+\frac{1}{  r^2} \,  p_\te^2\right)+\beta_0  r^2 }&\quad
 \displaystyle{
{H}^{\SStc} =\frac 12   \left(p_r^2
-\frac{1}{   r^2} \,  p_\te^2\right)+\beta_0  r^2 }\\[8pt]
 \displaystyle{\qquad\quad
+\frac{1}{  r^2}\left( \frac{\beta_1}{\cos^2\te}+\frac{\beta_2}{\sin^2 \te}  
\right)} &\quad
 \displaystyle{\qquad\quad
+\frac{1}{   r^2}\left( \frac{\beta_1}{\cosh^2\te}-\frac{\beta_2}{\sinh^2 \te}  
\right) }\\[8pt]
 \displaystyle{\qquad\ \,= \frac 12 \,p_r^2 
+\frac{1}{2   r^2}\, {C}_z+ \beta_0  r^2  } &\quad
 \displaystyle{\qquad\ \,= \frac 12 \,p_r^2 
-\frac{1}{2   r^2}\, {C}_z+ \beta_0 r^2  }\\[10pt]
 \displaystyle{ {C} =p_\te^2+ \frac{2\beta_1}
{\cos^2 \te}  +\frac{2 \beta_2}{\sin^2 \te }    }&\quad
 \displaystyle{ {C} =p_\te^2- \frac{2\beta_1}{
\cosh^2 \te }+\frac{2 \beta_2}{\sinh^2 \te }     }  \\[10pt]
 \displaystyle{ {I} =\left(\sin\te \,p_r+\frac{\cos\te}{  r}\,p_\te \right)^2
   }&\quad
 \displaystyle{  {I} =\left(\sinh\te \,p_r-\frac{\cosh\te}{  r}\,p_\te \right)^2
}\\[8pt]
 \displaystyle{ \qquad
+2\beta_0  r^2\sin^2\te +\frac{2\beta_2}{  r^2\sin^2\te}   }&\quad
 \displaystyle{ \qquad
+2\beta_0  r^2\sinh^2\te +\frac{2\beta_2}{  r^2\sinh^2\te}   }\\[14pt]

 \mbox {$\bullet$ Hyperbolic space ${\bf H}^2$}&\quad\mbox {$\bullet$ De
Sitter spacetime ${\bf dS}^{1+1}$}\\[4pt]
 \displaystyle{
{H}^{\SStc}_z=\frac 12   \left(p_r^2
+\frac{1}{ \sinh^2  r} \,  p_\te^2\right)+\beta_0\tanh^2 r }&\quad
 \displaystyle{
{H}^{\SStc}_z=\frac 12   \left(p_r^2
-\frac{1}{ \sinh^2  r} \,  p_\te^2\right)+\beta_0\tanh^2 r }\\[8pt]
 \displaystyle{\qquad\quad
+\frac{1}{\sinh^2  r}\left( \frac{\beta_1}{\cos^2\te}+\frac{\beta_2}{\sin^2 \te}  
\right)} &\quad
 \displaystyle{\qquad\quad
+\frac{1}{\sinh^2  r}\left( \frac{\beta_1}{\cosh^2\te}-\frac{\beta_2}{\sinh^2 \te}  
\right) }\\[8pt]
 \displaystyle{\qquad\ \,= \frac 12 \,p_r^2 
+\frac{1}{2\sinh^2  r}\, {C}_z+ \beta_0\tanh^2 r  } &\quad
 \displaystyle{\qquad\ \,= \frac 12 \,p_r^2 
-\frac{1}{2\sinh^2  r}\, {C}_z+ \beta_0\tanh^2 r  }\\[10pt]
 \displaystyle{ {C}_z=p_\te^2+ \frac{2\beta_1}
{\cos^2 \te}  +\frac{2 \beta_2}{\sin^2 \te }    }&\quad
 \displaystyle{ {C}_z=p_\te^2- \frac{2\beta_1}{
\cosh^2 \te }+\frac{2 \beta_2}{\sinh^2 \te }    }  \\[10pt]
 \displaystyle{ {I}_z=\left(\sin\te \,p_r+\frac{\cos\te}{\tanh r}\,p_\te \right)^2
   }&\quad
 \displaystyle{  {I}_z=\left(\sinh\te \,p_r-\frac{\cosh\te}{\tanh r}\,p_\te \right)^2
}\\[8pt]
 \displaystyle{ \qquad
+2\beta_0\tanh^2 r\sin^2\te +\frac{2\beta_2}{\tanh^2 r\sin^2\te}   }&\quad
 \displaystyle{ \qquad
+2\beta_0\tanh^2 r\sinh^2\te +\frac{2\beta_2}{\tanh^2 r\sinh^2\te}   } \\[8pt]
\hline
\end{array}
$$
\hfill}
\end{table}

\newpage


\epsfbox{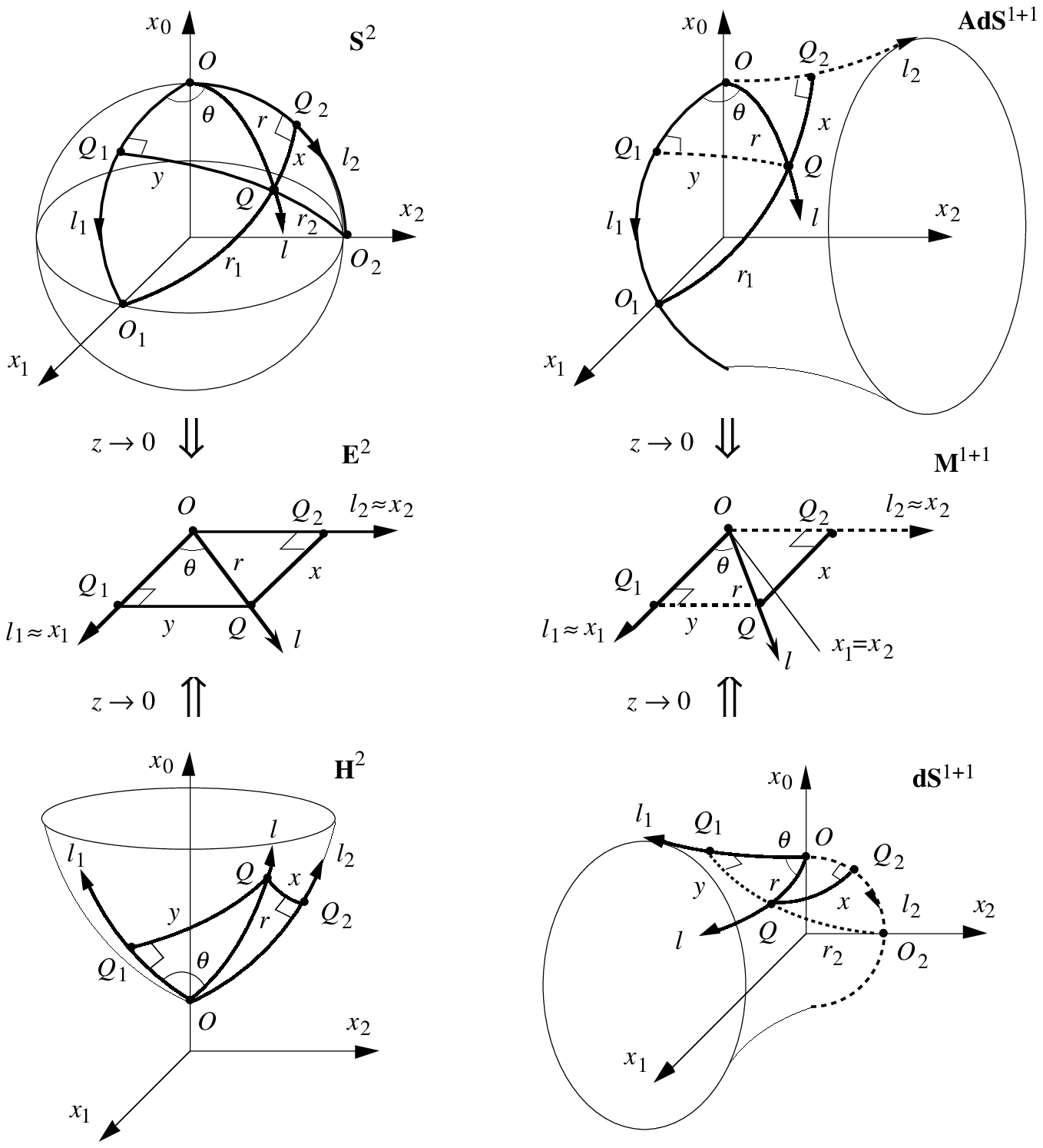}

\medskip
\medskip

\noindent
{{Figure 1.} Geometrical 
description of the  SW   potential  on the six  spaces of tables
\ref{table3} and \ref{table4}    in a 3D linear ambient space. In the three
spacetimes (right) space-like lines/distances are represented by dashed lines}



{\footnotesize



\begin{thebibliography}{99}


\bibitem{Dr}  Drinfel'd V G  1983  {\it Sov. Math. Dokl.}  {\bf 27}  68 


\bibitem{Dri}  Drinfel'd V G  1987 in 
{\it  Proc.  Int. Congress of
Math.  (Berkeley  1986)} ed    A V  Gleason  
(Providence, RI: American Mathematical Society) p 798

\bibitem{CP}   Chari V and  Pressley A  1994  {\it A Guide to Quantum Groups}
(Cambridge: Cambridge University Press)

\bibitem{majid}   Majid S  1995 {\it Foundations of Quantum
Group Theory}  (Cambridge: Cambridge University Press)

\bibitem{Sierra}  Gomez C,  Sierra G and   Ruiz-Altaba M 1996    {\it Quantum
Groups in Two-Dimensional Physics} (Cambridge: Cambridge University Press)



\bibitem{BR} Ballesteros A and Ragnisco O 1998 {\em J.
Phys. A: Math. Gen.} {\bf 31} 3791

\bibitem{Deform}
     Ballesteros A and Herranz F J    1999
{\it J. Phys. A: Math. Gen.} {\bf 32}  8851







\bibitem{photon}  Ballesteros A and  and Herranz F J 2000 
{\em Czech. J. Phys.} {\bf 50} 1239


\bibitem{photonb}  Ballesteros A and  and Herranz F J 2001 
{\em J. Nonlinear Math. Phys.} {\bf 8} 18

\bibitem{cluster}  Ballesteros A and  Ragnisco O 2003 
{\em J. Phys. A: Math. Gen.} {\bf 36} 10505


 \bibitem{CRMAngel}
     Ballesteros A,  Herranz F J, Musso F and Ragnisco O    2004
in {\em Superintegrability in Classical and Quantum  
Systems} {CRM Proceedings and Lecture Notes} {\bf 37} ed P Tempesta {\em et al}
(Providence, RI: American Mathematical Society) p 1
({\em Preprint} math-ph/0412067)

  



\bibitem{plb} Ballesteros A, Herranz F J and Ragnisco O 2005 {\em Phys.
Lett. B} {\bf 610} 107

  


\bibitem{Ohn}   Ohn C 1992   
{\it Lett. Math. Phys.} {\bf 25}  85








 

\bibitem{Fris}
Fris J,  Mandrosov V,  Smorodinsky Ya A,  Uhlir M  and  Winternitz P  1965
{\em Phys. Lett.} {\bf 16} 354 

 


\bibitem{Evansa}
Evans N  W  1990 {\em Phys. Lett. A} {\bf 147}  483

\bibitem{Evansb}
Evans N  W  1991 {\em J. Math. Phys.} {\bf 32}  3369

 
  
\bibitem{Groschea}
   Grosche  C,  Pogosyan G S   and  Sissakian A N 1995 {\em  Fortschr. Phys.}
{\bf 43} 453 



\bibitem{Per} Perelomov A M 1990 {\it Integrable Systems of Classical
Mechanics and Lie algebras}  (Berlin: Birkh\"auser)



\bibitem{groscheS2S3}
         Grosche C, Pogosyan G S and Sissakian A N  1995
{\it Fortschr. Phys. } {\bf 43}  523


\bibitem{KalninsH2}
         Kalnins E G,   Miller W Jr  and  Pogosyan G S   1997
{\it J. Math. Phys. } {\bf 38}  5416


 \bibitem{RS}
       Ra\~nada M F and  Santander M   1999
{\it J. Math. Phys. } {\bf 40}  5026

 \bibitem{PogosClass1}
         Kalnins E G,   Miller W Jr  and  Pogosyan G S   2000
{\it J. Phys. A: Math. Gen.} {\bf 33}  6791


 \bibitem{PogosClass2}
         Kalnins E G, Kress J M,   Pogosyan G S  and
 Miller W Jr   2001
{\it J. Phys. A: Math. Gen.} {\bf 34}  4705




\bibitem {Yaglom} 
Yaglom I M 1979
{\it A Simple Non-Euclidean Geometry and its Physical Basis}   
(New York: Springer)   



\bibitem{Trigo}   Herranz F J, Ortega R and Santander M   2000
{\it J. Phys. A: Math. Gen.} {\bf 33}  4525


\bibitem{Conf}   Herranz F J and Santander M   2002
{\it J. Phys. A: Math. Gen.} {\bf 35}  6601


\bibitem {Doub} 
 Doubrovine B,  Novikov S and  Fomenko A  1982 
{\it G\'eom\'etrie Contemporaine, M\'ethodes et Applications}  First Part
(Moscow: MIR)  




 

 
\bibitem{Higgs}
     Higgs P W   1979
{\it J. Phys. A: Math. Gen.} {\bf 12}  309

\bibitem{Leemon}
     Leemon H I  1979
{\it J. Phys. A: Math. Gen.} {\bf 12}  489



\bibitem{Evans}
Evans N  W  1990 {\em Phys. Rev. A} {\bf 41}  5666


 
  \bibitem{Schrodingerdual}
        Kalnins E G,   Miller W Jr  and  Pogosyan G S  2000
{\it J. Math. Phys. } {\bf 41}  2629


  \bibitem{Schrodingerdualb}
        Nersessian A and Pogosyan G S  2001
  {\it Phys. Rev. A} {\bf 63} 020103(R)



 \bibitem{Schrodingerdualc}
        Kalnins E G,   Miller W Jr  and  Pogosyan G S  2002
{\it   Phys. Atom. Nuclei} {\bf 65}  1086




  \bibitem{Schrodinger}
   Schr\"odinger  E     1940
{\it Proc. R. Ir. Acad. A} {\bf 46}  9

  
 


\bibitem{VulpiLett}
   Herranz F J,  Ballesteros A,  Santander M and Sanz-Gil T    2003
{\it J. Phys. A: Math. Gen.} {\bf 36}  L93

 

\bibitem{CRMVulpi}
   Herranz F J,  Ballesteros A,  Santander M and Sanz-Gil T    2004
in {\em Superintegrability in Classical and Quantum  
Systems}
{CRM Proceedings and Lecture Notes} {\bf 37} ed P Tempesta {\em et al}
(Providence, RI: American Mathematical Society) p 75
({\em Preprint} math-ph/0501035)
 

 \bibitem{ran}
    Ra\~nada  M  F  and   Santander M     2002
{\it Rep. Math. Phys.} {\bf 49}  335


 \bibitem{ran1}
    Ra\~nada  M  F  and   Santander M     2002
{\it J. Math. Phys.} {\bf 43}  431

 \bibitem{ran2}
    Ra\~nada  M  F  and   Santander M     2003
{\it J. Math. Phys.} {\bf 44}  2149



  \bibitem{ranran}
    Ra\~nada  M  F  and   Santander M  and  Sanz-Gil T  2002 
in {\it Classical and Quantum Integrability}
Banach Center Publications {\bf 59} (Warsaw: Polish Acad.  Sci.) p 243



 \bibitem{KKWa}
         Kalnins E G, Kress J M and Winternitz P 2002
{\it J. Math. Phys.} {\bf 43}  970

 \bibitem{KKWb}
         Kalnins E G, Kress J M, 
 Miller W Jr  and Winternitz P 2003
{\it J. Math. Phys.} {\bf 44}  5811



\end{thebibliography}


}
\end{document}